\documentclass[10pt]{article}

\usepackage[a4paper,margin=0.7in]{geometry}
\usepackage[english]{babel}
\usepackage{float}
\usepackage[dvipsnames]{xcolor}
\usepackage{graphicx}
\usepackage{amsmath, amsthm, amssymb}
\usepackage{amsfonts}
\usepackage{romannum}
\usepackage{tikz}
\usepackage{titlesec}
\usepackage{abstract}
\usepackage[numbers,sort&compress]{natbib}
\usepackage{caption}
\usepackage{subcaption}
\usepackage{mathrsfs} 
\usepackage{upgreek}
\usepackage[T1]{fontenc}
\usepackage{lmodern}

\usepackage[
    colorlinks=true,
    linkcolor=red,
    filecolor=blue,
    urlcolor=blue,
    citecolor=blue,
    breaklinks=true
]{hyperref}

\newcommand{\orcid}[1]{%
  \,\href{https://orcid.org/#1}{\raisebox{-0.4pt}{\includegraphics[width=8pt]{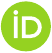}}}%
}

\renewcommand{\thesection}{\Roman{section}}
\renewcommand{\thesubsection}{\Roman{section}.\Alph{subsection}}

\titleformat{\section}
  {\large\bfseries\fontsize{10}{12}\selectfont\MakeUppercase}
  {\thesection.}{0.5em}{}

\titleformat{\subsection}
  {\normalsize\bfseries}
  {\thesubsection.}{0.5em}{}

\title{%
\vspace{-1.5em}
\textbf{\fontsize{16pt}{16pt}\selectfont\bfseries Scanless quantum Fourier-transform mid-infrared spectroscopy for solids and surface analysis}
\vspace{-0.5em}
}

\author{%
\parbox{0.96\textwidth}{\centering
\normalsize
Paul Gattinger\orcid{0000-0002-0120-9100}$^{1,*}$\quad
Michela Conti\orcid{0009-0009-7675-1445}$^{1}$\quad
Franziska Dietz\orcid{0009-0009-9749-9162}$^{1}$\\[0.25em]
Andrei N. Halangescu Moldovan\orcid{0009-0008-8736-8436}$^{1}$\quad
Andreas W. Schell\orcid{0000-0003-0849-9558}$^{2}$\quad
Markus Brandstetter\orcid{0000-0002-8679-8097}$^{1}$\quad
Ivan Zorin\orcid{0000-0002-2089-5716}$^{1,*}$\\[0.8em]
\footnotesize\itshape
$^{1}$Research Center for Non-Destructive Testing, Science Park 2, Altenberger Str. 69, 4040 Linz, Austria\\
$^{2}$Division of Light-Matter-Interaction, Johannes Kepler University, Altenberger Str. 69, 4040 Linz, Austria\\[0.6em]
\footnotesize\normalfont
$^{*}$Corresponding authors: \href{mailto:paul.gattinger@recendt.at}{paul.gattinger@recendt.at} and \href{mailto:ivan.zorin@recendt.at}{ivan.zorin@recendt.at}\\[0.6em]
\footnotesize
\today
}
}
\date{}

\begin{document}
\maketitle
\vspace{-2em}

\begin{onecolabstract} 
Quantum infrared (IR) spectroscopy is a novel technique based on nonlinear interferometry and quantum sensing with undetected photons. In this approach, spectrally far-separated correlated photon pairs are used for detection and probing, respectively. Hence, the probing wavelength domain (mid-IR) is substantially different from the detection spectral range (near-IR), which enables shot-noise-limited and cost-effective sensing schemes attractive for applied metrology.
In this paper, we implement and employ a custom-built scanless quantum Fourier-transform mid-IR (sQFTIR) microspectrometer for chemical characterization and analysis of solids and interfaces. 
The sQFTIR system employs spectral-domain acquisition, without the need to scan the optical delay. This results in enhanced sensitivity and short measurement times and therefore enables rapid spectral acquisitions in 100~ms spanning a spectral band from 3000~cm$^{-1}$ -- 2400~cm$^{-1}$. The capabilities of our approach are demonstrated for hyperspectral mid-IR imaging of dried pharmaceuticals (insulin on a gold surface) and investigation of a security feature on a banknote. Furthermore, we demonstrate hyperspectral imaging and spectroscopy through a turbid ceramic medium.
\end{onecolabstract}
\vspace{0.5em}
\saythanks          

\section{Introduction}

Mid-infrared (mid-IR) spectroscopy is a powerful analytical tool to probe molecular bonds in all aggregate states of matter~\cite{griffiths2007fourier,Saptari2003}. While the classical Fourier-transform IR spectrometer (FTIR) has long served as the workhorse of IR spectroscopy, newer laser-based techniques have entered the field in the past decades. Mid-IR spectroscopic techniques can be largely divided into direct and indirect techniques. Direct techniques measure the absorption of mid-IR light from classical thermal sources, mid-IR quantum cascade lasers~\cite{kosterev2002,Kosterev2008,schwaighofer2017} or supercontinuum laser sources~\cite{Zorin:22}. In contrast, indirect techniques rely on secondary effects such as photothermal~\cite{zhang2026,zhu2026} or photoacoustic response~\cite{patimisco2014}. A key limiting factor of direct techniques is the low sensitivity of available detector platforms such as mercury cadmium telluride (MCT)~\cite{Rogalski2010}, inhibiting shot-noise-limited measurements. Other aspects constraining both indirect and direct techniques are high source powers, imposing a risk of destroying delicate samples, high noise of coherent sources, and high costs of mid-IR components. 

In recent years, sensing with undetected photons, introduced by Lemos et al.~\cite{lemos2014}, has gained increasing relevance and interest in mid-IR spectroscopy. The technique uses correlated (energy- and time-entangled) photon pairs, facilitating the separation of the probing and the detection spectral ranges. Thereby, mid-IR absorption, phase changes, and scattering can be measured in the near-IR regime using efficient shot-noise-limited Si-based detectors and less than a nanowatt of power impinging on the sample. 
The first reports of mid-IR spectroscopy with undetected photons, referred to as quantum mid-IR spectroscopy, were performed in the 3.8--4.5~$\mu$m range and focused on sensing of CO$_2$~\cite{Kalashnikov2016,Paterova2017}. In the following years, quantum FTIR (QFTIR) spectroscopy for gas and solid-phase measurements has been introduced~\cite{Lindner2020,Lindner2022}; at the same time, the spectral coverage was pushed further into the mid-IR fingerprint region~\cite{Mukai2022,Paterova2022}, while alternative acquisition schemes were investigated~\cite{Kaufmann:22}. Furthermore, relevant application-oriented techniques such as attenuated total reflection spectroscopy~\cite{Kurita2025} or rapid material identification~\cite{Sherwani2026} were demonstrated. Enhancement techniques such as ultra-broadband spectral bandwidth~\cite{Tashima2024} or pump-enhancement~\cite{Lindner2023} were showcased, indicating a promising path for applied quantum IR spectroscopy. Moreover, the development of hyperspectral microscopic quantum mid-IR imaging~\cite{Paterova2020,Kviatkovsky2020,Suryana2025,Placke2026,gattinger2026preprint} has the potential to make IR hyperspectral microscopic imaging more accessible and significantly more affordable for applied and biomedical research.

Here, we implement and deploy a custom-built mid-IR hyperspectral microscope based on scanless QFTIR spectroscopy, operating in the wavelength band between 3000 -- 2400~cm$^{-1}$, for applied non-destructive investigation of surfaces and interfaces of solids. Our sQFTIR instrument performs spatial mapping of the sample with measurement times of 100~ms per hyperspectral pixel, thus rendering it competitive with state-of-the-art hyperspectral mid-IR techniques. In the first part of the manuscript, principles of the technique along with signal acquisition and processing are introduced.
In the experimental section, we showcase quantum mid-IR hyperspectral imaging and microspectroscopy of diverse and relevant samples: a dried pharmaceutical on a gold surface, a banknote security feature, and spectroscopy through turbid media.

\section{Scanless quantum FTIR}
The operational principle of sQFTIR is based on sensing with undetected photons. This technique exploits the process of spontaneous parametric down-conversion (SPDC), where pump photons with frequency $\omega_p$ are converted to spectrally non-degenerate energy- and time-entangled photons, called signal ($\omega_s$) and idler ($\omega_i$). Typically, periodically poled nonlinear crystals with poling period $\Lambda$ are used as SPDC sources~\cite{Chekhova2016,Boyd2020}. The energy entanglement, i.e., the correlations within the bi-photon, can be expressed as $\omega_p = \omega_s + \omega_i$, the momentum conservation is expressed via the quasi-phase-matching condition $\Delta k = k_p - k_s - k_i - 2\pi/\Lambda$, where $\Delta k$ is the wavevector mismatch (used to design the SPDC profile). SPDC can be considered a quantum process since the signal ($\omega_s$, near-IR) and idler ($\omega_i$, mid-IR) modes emerge spontaneously from the vacuum state. The idler photons are usually separated from the signal and pump photons in order to probe a sample in transmission or reflection. Pump, signal, and idler photons are then overlapped in a second SPDC process that again possesses a certain probability of pair generation. It has to be noted that, in the low-gain regime considered here, difference frequency generation can be neglected. Due to path-indistinguishability~\cite{Zou1991,Wang1991,Mandel1991}, the joint (signal-idler) generation probability amplitudes in the first and second SPDC events start to interfere. The interference is recorded in the signal domain; idler photons remain undetected. In a simple single-mode approximation, the count rate of the signal photons $\rho_s$ at the detector is proportional to~\cite{BarretoLemos:22}:
\begin{equation}
    \rho_s \propto \frac{1}{2} \left(1+|t_i|\cos(\gamma + \Delta \phi)\right), 
    \label{eq:rate}
\end{equation}
where $|t_i|$ is the transmission amplitude of the sample, $\gamma$ is the phase shift introduced by the sample and $\Delta \phi$ is the phase mismatch between pump, signal and idler $\Delta \phi = \phi_p - \phi_s - \phi_i$. Equation~\ref{eq:rate} indicates that the complex transmission amplitude of a sample inserted in the probing (idler) beam can be inferred from the signal interference; the idler photons $\omega_i$ can be disregarded since the desired information is accessible in the signal photon range.

In the experimental realization of quantum IR spectroscopy with undetected photons, a periodically poled potassium titanyl phosphate crystal (ppKTP) with a poling period $\Lambda$ was used as nonlinear material for SPDC. Instead of an arrangement of two nonlinear crystals, a double-pass configuration (forward and backward) through the same crystal was implemented.
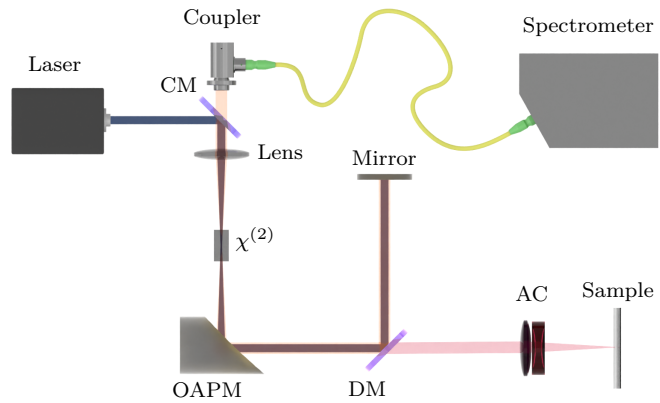
\begin{figure}[!b]
    \centering
    \begin{tikzpicture}
        \node[
            anchor=south west,
            inner sep=0
        ] (image) at (0,0) {%
            \includegraphics[width=0.62\textwidth]{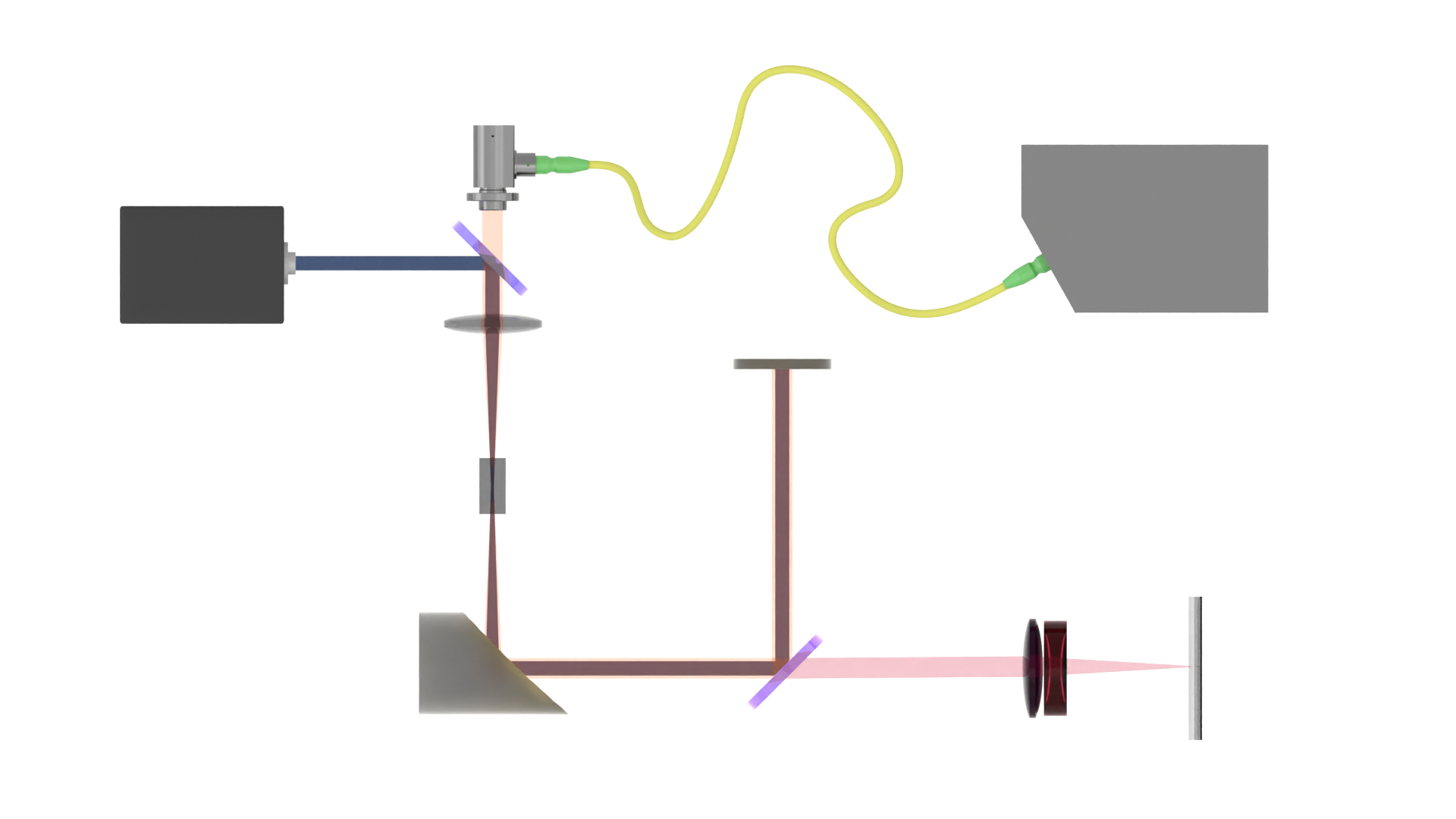}%
        };
        \begin{scope}
        [
            x={(image.south east)},
            y={(image.north west)},
            every node/.style={
                font=\small,
                text=black,
                inner sep=1.5pt
            }
        ]
        \draw (0.537,0.596) node[]{\color{black}\footnotesize Mirror};
        \draw (0.822,0.306) node[]{\color{black}\footnotesize Sample};
        \draw (0.717,0.302) node[]{\color{black}\footnotesize AC};
        \draw (0.516,0.096) node[]{\color{black}\footnotesize DM};
        \draw (0.321,0.093) node[]{\color{black}\footnotesize OAPM};
        \draw (0.38,0.42) node[]{\color{black}\footnotesize $\chi^{(2)}$};
        \draw (0.41,0.61) node[]{\color{black}\footnotesize Lens};
        \draw (0.286,0.754) node[]{\color{black}\footnotesize CM};
        \draw (0.339,0.898) node[]{\color{black}\footnotesize Coupler};
        \draw (0.135,0.800) node[]{\color{black}\footnotesize Laser};
        \draw (0.786,0.881) node[]{\color{black}\footnotesize Spectrometer};
        \end{scope}
    \end{tikzpicture}
    \caption{Experimental setup: the sQFTIR hyperspectral imaging system based on a nonlinear interferometer with a Michelson topology. A pump beam (660~nm continuous-wave laser) is sent into the system through a cold mirror (CM) and focused into a ppKTP crystal type-0 phase-matched for non-degenerate SPDC. In the first pump-pass, the nonlinear crystal produces idler-signal photon pairs through SPDC. The idler, signal and pump beams are collimated by an off-axis parabolic mirror (OAPM) and sent to a dichroic mirror (DM) that reflects pump and signal photons while transmitting the idler. An achromatic doublet focuses the idler photons on the sample, which is mounted on a x-y translational stage. The signal and pump photons are backreflected by a silver mirror. All photons (pump, signal, and backscattered idler) propagate back to the nonlinear crystal where second SPDC process happens again. If the photon paths are indistinguishable (overlapped), bi-photon interference can be observed after the second pass. The signal photons are transmitted though the cold mirror and coupled into the fiber to a commercial spectrometer.}
    \label{fig:system}
\end{figure}
The scheme of the experimental sQFTIR modality is shown in Fig.~\ref{fig:system}. A 660 nm pump laser (Cobolt Flamenco, 500~mW) is launched into the nonlinear interferometer (Michelson-type~\cite{Chekhova2016}) via a cold mirror (CM, M254C45, Thorlabs) and an achromatic (AC) doublet ($f=75$~mm, AC254-075-B-ML, Thorlabs) focuses it into the ppKTP crystal ($l=2.55$~mm, $\Lambda = 20.45$~\textmu m, $T=20$°C, SLF LaserFabriken,) with a focusing parameter of 1.42. The emerging pump, signal and idler photons are collimated via a silver coated off-axis parabolic mirror (OAPM, f=75~mm). A custom dichroic mirror (DM) reflects the pump and signal photons towards a fixed mirror and transmits the idler photons towards an achromatic doublet ($f=50$~mm, AC254-050-E, Thorlabs), focusing them onto the sample. The sample is placed on an x-y stage for spatial mapping. Reflected pump, signal, and idler photons overlap a second time in the ppKTP crystal, where the second SPDC process is triggered (without seeding of difference frequency generation). The signal photons are then transmitted through the CM and coupled into the spectrometer (Ocean Optics QE Pro, customized), where spectral bi-photon interference is measured (without scanning of the reference arm). Since the transmission amplitude of the sample is directly proportional to the fringe visibility $V = (\rho_{\mathrm{s,max}}-\rho_{\mathrm{s,min}})/(\rho_{\mathrm{s,max}}+\rho_{\mathrm{s,min}})$~\cite{Kalashnikov2016}, post-processing is needed to infer the spectral visibility, and thus the transmission in each spectral component.

In the post-processing steps~\cite{gattinger2026preprint}, an inverse Fourier transform is applied to the spectral interferogram, and the coherence function is located and isolated from the DC component (the baseband is Fourier-shifted to 0 frequency); the distance of the coherence burst from the DC component (i.e., the group delay $\Delta z$) defines the spectral resolution $\delta \nu$. Afterwards, a symmetric apodization function (Blackman–Harris, 3-term) is applied to the isolated burst (similar to a center burst in classical FTIR). In the next step, zero padding is applied, and the signal is Fourier-transformed. By taking the magnitude of the derived complex envelope, the mid-IR spectrum is recovered without scanning of the group delay.

\section{Measurements of pharmaceutical ingredients on a gold surface}
At-line chemical analysis of surfaces is of particular practical relevance in pharmaceutical production. The verification of equipment and surface cleanliness (vessels, pipes) and the detection of residual pharmaceutical compounds on product-contact surfaces are integral parts of pharmaceutical manufacturing and are required to prevent cross-contamination between production cycles. Conventional methods rely on swab or rinse sampling followed by laboratory-based analysis, while direct FTIR-based approaches provide rapid measurements but are localized and averaged over a spot produced by an extended thermal source~\cite{SARWAR2022100130}.
In this context, sQFTIR hyperspectral imaging is a promising alternative that enables spatially resolved identification in the mid-IR at reasonable costs with a high resolution over an extended surface area.
In order to emphasize the applicability of the sQFTIR-based hyperspectral microscope for at-line chemical analysis of surfaces, a pharmaceutical sample was mapped.

The insulin sample was prepared by drop-casting 10~\textmu L of a commercially available recombinant human insulin solution (nominal concentration 10~mg/mL; I9278, Sigma-Aldrich) onto a planar gold surface, (PF10-03-M03, Thorlabs) yielding a nominal absolute mass of 100~\textmu g of insulin.
The droplet was allowed to evaporate completely under ambient conditions.
This unconstrained drying process induced a characteristic coffee-ring effect, leading to macroscopic crystallization and localized material accumulation at the droplet perimeter.
The dried insulin drop was then raster scanned (397 $\times$ 200 pixels) using the sQFTIR setup shown in Fig.~\ref{fig:system}. The integration time used per spectrum (i.e., spatio-spectral pixel) was 100~ms, and the step size in x- and y-directions was 5~\textmu m. The experimentally achieved spatial resolution was determined to be 12.3~\textmu m~\cite{gattinger2026preprint} and the spectral resolution was set to 20~cm$^{-1}$, as defined by the selected group delay $\Delta z$. The sQFTIR post-processing was applied pixel-wise, and absorbance spectra were calculated for each pixel according to $A=-\log_{10}(I/I_0)$, with $I$ being the transflected intensity from the sample (transflection amplitude) and $I_0$ being the reflected intensity from the gold coating. For a scattering sample, the absorbance must be interpreted as apparent absorbance, since scattering may introduce a nonlinear baseline. 

\begin{figure}[!b]
    \centering

    \begin{subfigure}[c]{0.29\textwidth}
        \centering
        \includegraphics[width=\linewidth]{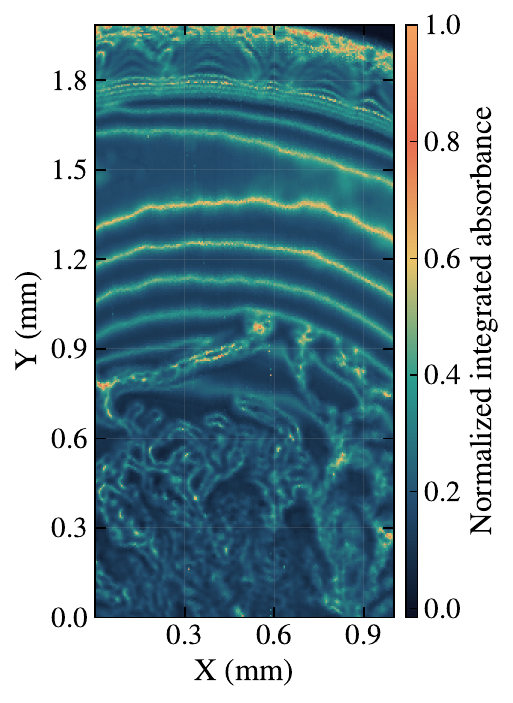}
        \caption{Spectral image of a dried insulin drop}
        \label{fig:insulin_image}
    \end{subfigure}
    \hspace{0.2cm}
    \begin{subfigure}[c]{0.52\textwidth}
        \centering
        \includegraphics[width=\linewidth]{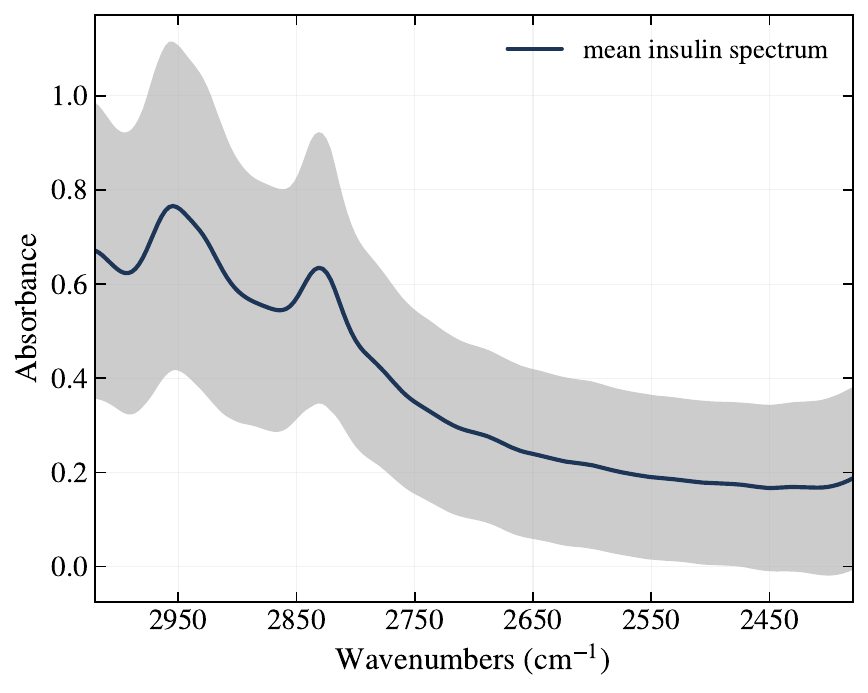}
        \caption{Average mid-IR absorbance spectrum and standard deviation}
        \label{fig:insulin_spectra}
    \end{subfigure}
    \hspace{0.2cm}
    \caption{Hyperspectral imaging of a dried insulin drop on a gold surface. (a) Absorbance image integrated in the band between 2800-2850~cm$^{-1}$ (b) Average mid-IR absorbance spectrum taken from the hyperspectral cube. The gold mirror in the top right of the spectral image was used as a background for absorbance calculation. The standard deviation of the absorbance spectra is indicated as a grey area.}
    \label{fig:insulin}
\end{figure}

Figure~\ref{fig:insulin_image} shows a normalized integrated absorbance image obtained by integrating over the spectral band from 2800-2850~cm$^{-1}$. The part of the gold mirror that was used to calculate absorbance is visible in the top right corner of the image. Next to ripples presumably coming from the coffee-ring effect, insulin crystals are visible in the center and at the border of the drop. Additionally, thin film interference manifests on the outer border of the drop (top of the image). Variations in film thickness and thin-film interference lead to substantial variability in the absorbance spectra. The mean of the absorbance spectra is shown in Fig.~\ref{fig:insulin_spectra} and the standard deviation is indicated as a grey shaded area. The two distinctive absorbance bands correspond well to literature insulin spectra~\cite{Delbeck2021}.

\section{Hyperspectral imaging of a banknote security hologram}
The Europa portrait feature on the Europa-series 20€ is a sophisticated foil-over-window diffractive optical element. A visible makro image of a banknote with serial number EM2984824492 was recorded with a Thorlabs Telesto OCT system and is shown in Fig.~\ref{fig:vis_europa}. The eye area of Europa was mapped with the sQFTIR microscope (300 $\times$ 131 pixels) with a 10~\textmu m step size and 100~ms integration time per sQFTIR spectrum. The spectral resolution was set to 25~cm$^{-1}$. Two normalized integrated absorbance images at wavelength bands between 2940--2970~cm$^{-1}$ and 2990--3000~cm$^{-1}$ are shown in Fig.~\ref{fig:europa_image}. The background measurement needed for the calculation of absorbance spectra was obtained by a separate measurement of a gold mirror. Clear differences between the two images can be seen in the form of diffuse shadows on top of the eye area in the top image of Fig.~\ref{fig:europa_image}.

\newlength{\panelheight}
\setlength{\panelheight}{5.75cm}
\begin{figure}[htb]
    \centering

    \begin{subfigure}[t]{0.175\textwidth}
        \centering
        \begin{minipage}[t][\panelheight][t]{\linewidth}
            \centering
            \vspace{2mm}
            \setlength{\fboxsep}{0pt}
            \setlength{\fboxrule}{0.7pt}
            \fbox{%
                \includegraphics[
                    width=\linewidth
                ]{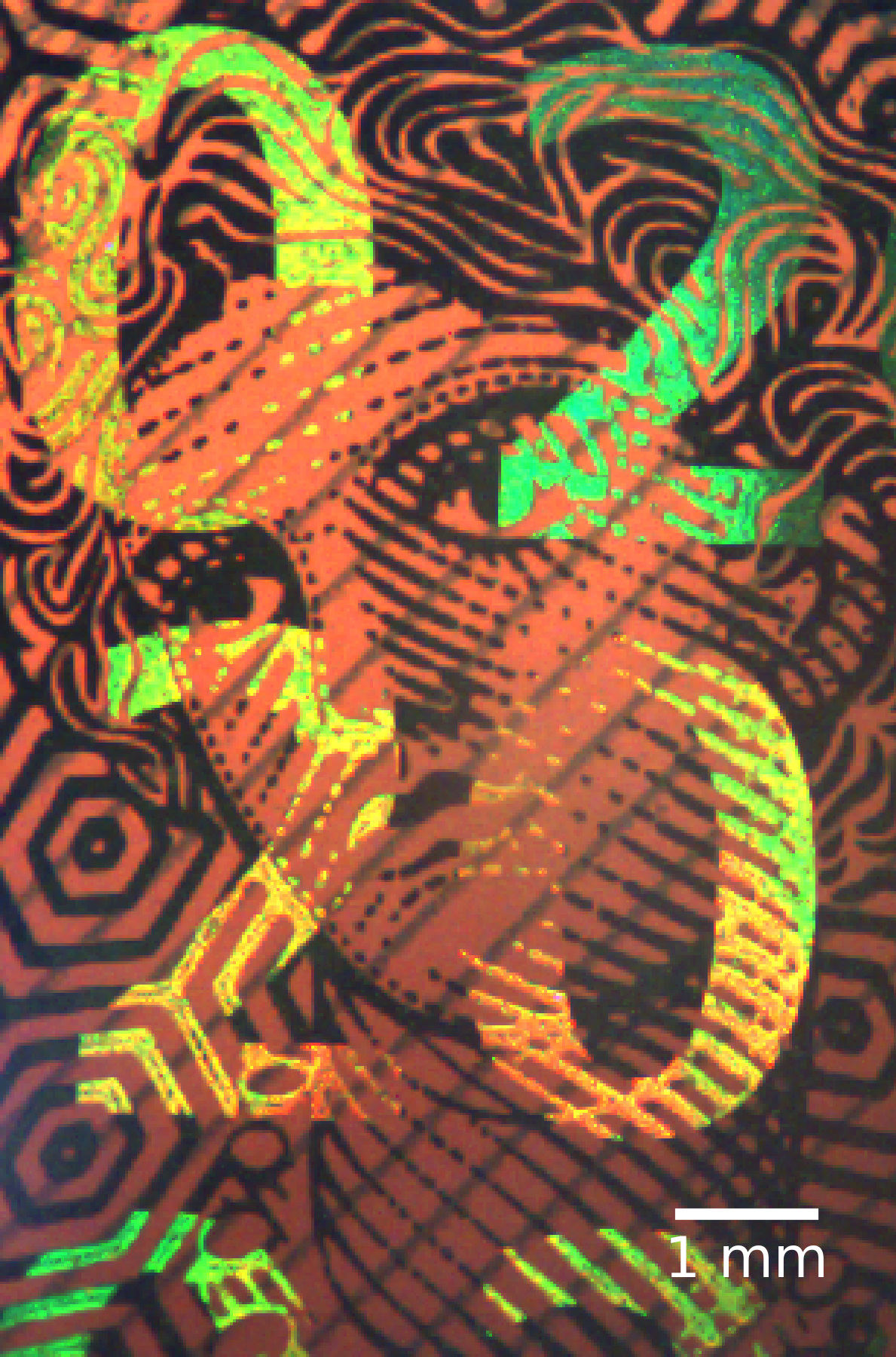}%
            }
        \end{minipage}

        \caption{Security feature of a banknote}
        \label{fig:vis_europa}
    \end{subfigure}
    \hspace{0.1cm}
    %
    \begin{subfigure}[t]{0.39\textwidth}
        \centering
        \begin{minipage}[t][\panelheight][t]{\linewidth}
            \centering
            \vspace{0pt}
            \includegraphics[
                width=\linewidth
            ]{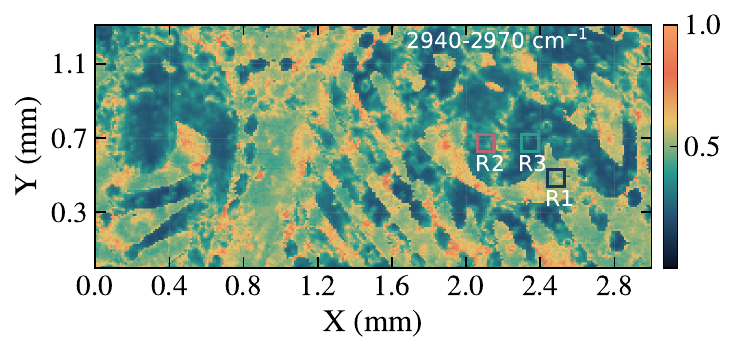}

            \vspace{-0.7cm}

            \includegraphics[
                width=\linewidth
            ]{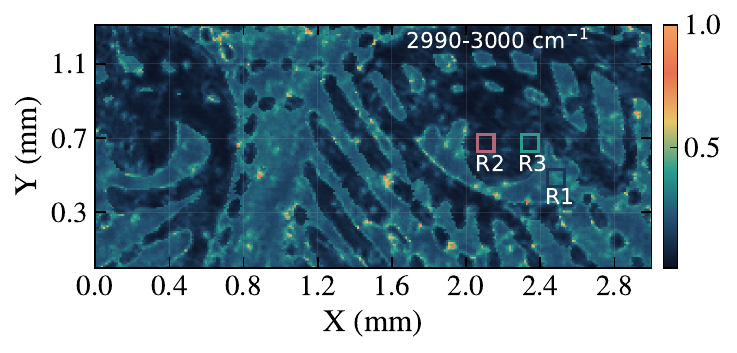}
        \end{minipage}

        \caption{Spectral images}
        \label{fig:europa_image}
    \end{subfigure}
    \hspace{0.05cm}
    %
    \begin{subfigure}[t]{0.33\textwidth}
        \centering
        \begin{minipage}[t][\panelheight][t]{\linewidth}
            \centering
            \vspace{1mm}
            \includegraphics[
                width=\linewidth
            ]{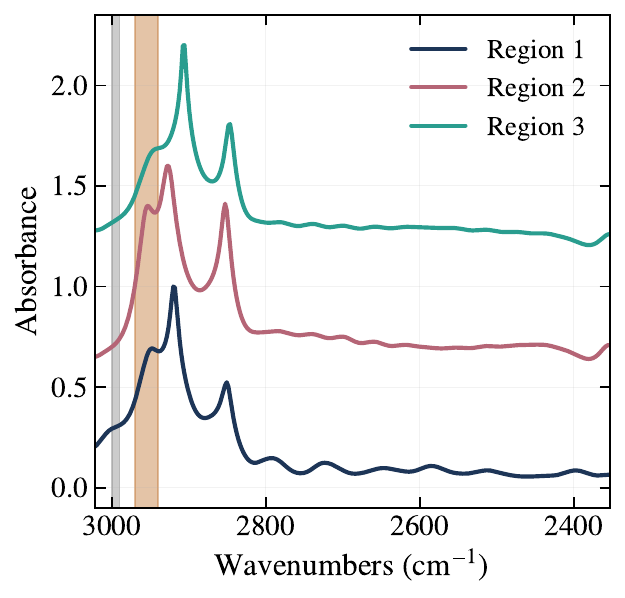}
        \end{minipage}

        \caption{Averaged spectra}
        \label{fig:europa_spectra}
    \end{subfigure}

    \caption{%
        (a) Visual image of a security feature on a 20€ banknote.
        (b) Top: spectral image integrated from 2940--2970~cm$^{-1}$;
        bottom: spectral image integrated from 2990--3000~cm$^{-1}$.
        (c) Averaged spectra extracted from different regions color coded in (b).
    }
    \label{fig:combined}
\end{figure}

Additionally, thin-film interference is a pronounced feature of the top image. There are three colored rectangles (10 $\times 10$ pixels) in the spectral images that indicate areas where the absorbance spectra shown in Fig.~\ref{fig:europa_spectra} have been extracted; the spectra are offset for better visibility. There are clear differences between the three regions. The most dominant difference is the band at 2910~cm$^{-1}$ in region 3, which is shifted towards smaller wavenumbers in the other two regions. Furthermore, the shoulder at 2950~cm$^{-1}$ is less pronounced in the averaged absorbance spectrum of region 3. Thin film interference manifests as fringes in the averaged absorbance spectrum of region 1. This feature can be used to estimate the thickness of the foil in this region according to $d={N}/{2n(\nu_1-\nu_2)}$, where $N$ is the number of averaged fringes and $\nu_{1,2}$ is the wavenumber of the first and the last selected fringe, respectively. Assuming a refractive index of 1.5, the resulting film thickness is estimated to be 48~\textmu m.

The obtained spectra cannot be unambiguously identified and assigned to a specific polymer; based on the positions and shapes of the C–H stretching bands, we suppose the composition of the transparent polymer carrier of the holographic foil strip to be made of polystyrene or polyethylene or a combination thereof (the CH band is insufficient to distinguish PE from PS confidently). Nevertheless, the sample is a complex structure composed of different layers (as seen in the mid-IR thin-film interference), including presumably protective coating and adhesive interfaces. Therefore, the measured spectra represent a superposition of the spectral contributions from the individual layers rather than the response of a single homogeneous material.

\section{Quantum infrared spectroscopy through a turbid medium}
\begin{figure}[!t]
    \centering
    \begin{subfigure}[t]{0.86\textwidth}
        \centering
        \begin{tikzpicture}
            \node[
                anchor=south west,
                inner sep=0
            ] (image) at (0,0) {%
                \includegraphics[width=0.55\linewidth]{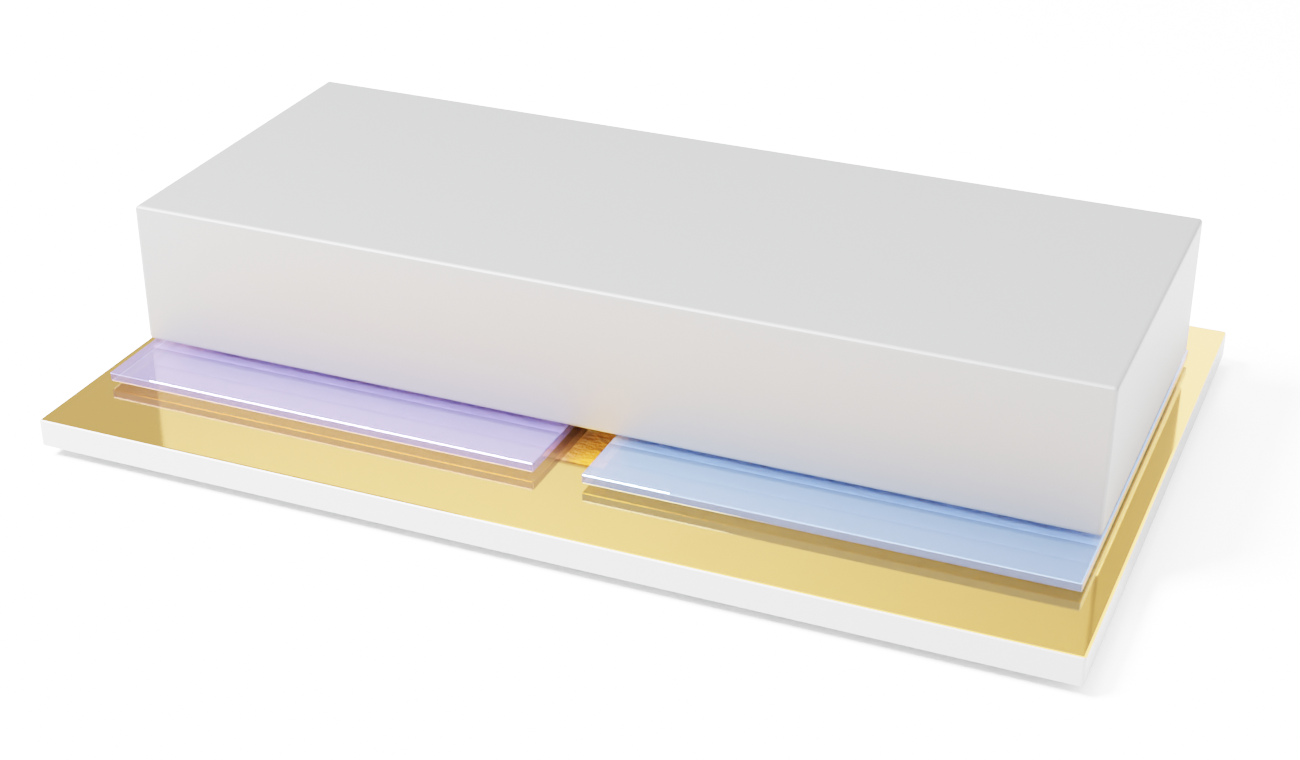}%
            };

            \begin{scope}[
                x={(image.south east)},
                y={(image.north west)},
                every node/.style={
                    font=\small,
                    text=black,
                    inner sep=1.5pt
                }
            ]
                \draw[stealth-stealth] (0.925,0.704) -- (0.914,0.533);
                \draw[stealth-]  (0.774,0.281) -- (0.925,0.165);
                \draw[stealth-]  (0.207,0.486) -- (0.091,0.247);
                \draw[stealth-]  (0.237,0.802) -- (0.130,0.923);
                \draw (0.99,0.6150) node[]{\color{black}\footnotesize 300~\textmu m};
                \draw (0.958,0.135) node[]{\color{black}\footnotesize PET};
                \draw (0.083,0.196) node[]{\color{black}\footnotesize PP};
                \draw (0.07,0.940) node[]{\color{black}\footnotesize Al\textsubscript{2}O\textsubscript{3}};
            \end{scope}
        \end{tikzpicture}
        \caption{Sample structure}
        \label{fig:turbid_sample}
    \end{subfigure}
    \vspace{0.4cm}
    \begin{subfigure}[t]{0.285\textwidth}
        \centering
        \includegraphics[width=\linewidth]{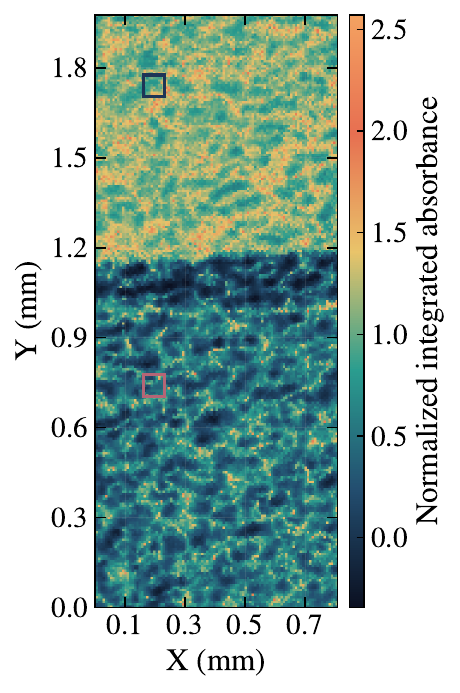}
        \caption{Spectral image}
        \label{fig:turbid_image}
    \end{subfigure}
    \hspace{0.2cm}
    \begin{subfigure}[t]{0.53\textwidth}
        \centering
        \includegraphics[width=\linewidth]{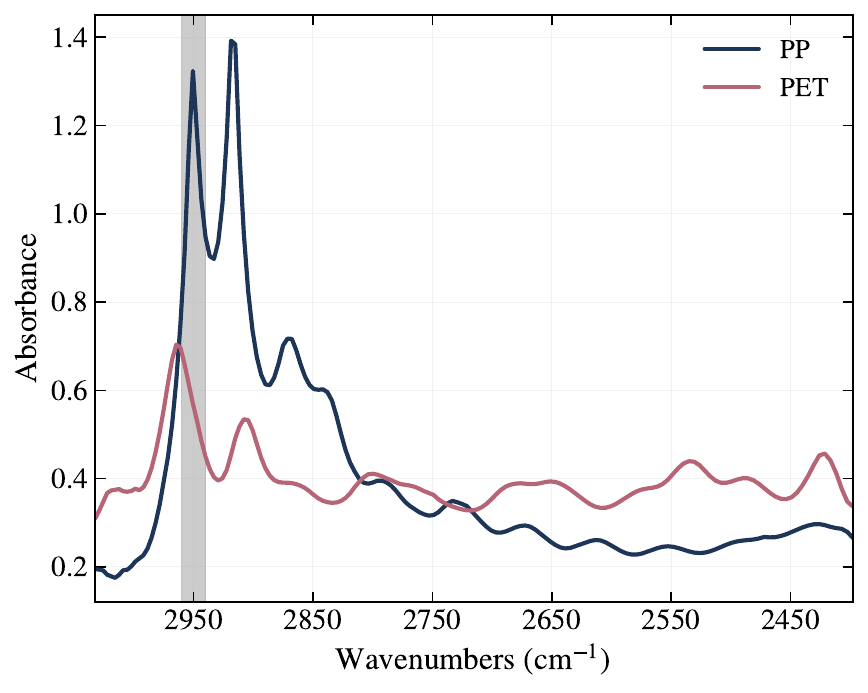}
        \caption{Extracted spectra}
        \label{fig:turbid_spectra}
    \end{subfigure}
    \caption{Microspectroscopy and hyperspectral imaging through a turbid medium: 
    (a) Sample structure consisting of an Al\textsubscript{2}O\textsubscript{3} wafer and polymer films (PP and PET) above a gold mirror. 
    (b) Spectral image of PP and PET underneath a turbid medium; the pixel values are the normalized integrated absorbance in the band between 2940--2960~cm$^{-1}$. 
    (c) Mean spectra of PP and PET extracted from the hyperspectral dataset; the areas used for averaging are indicated with colored rectangles in (b).}
    \label{fig:turbid}
\end{figure}
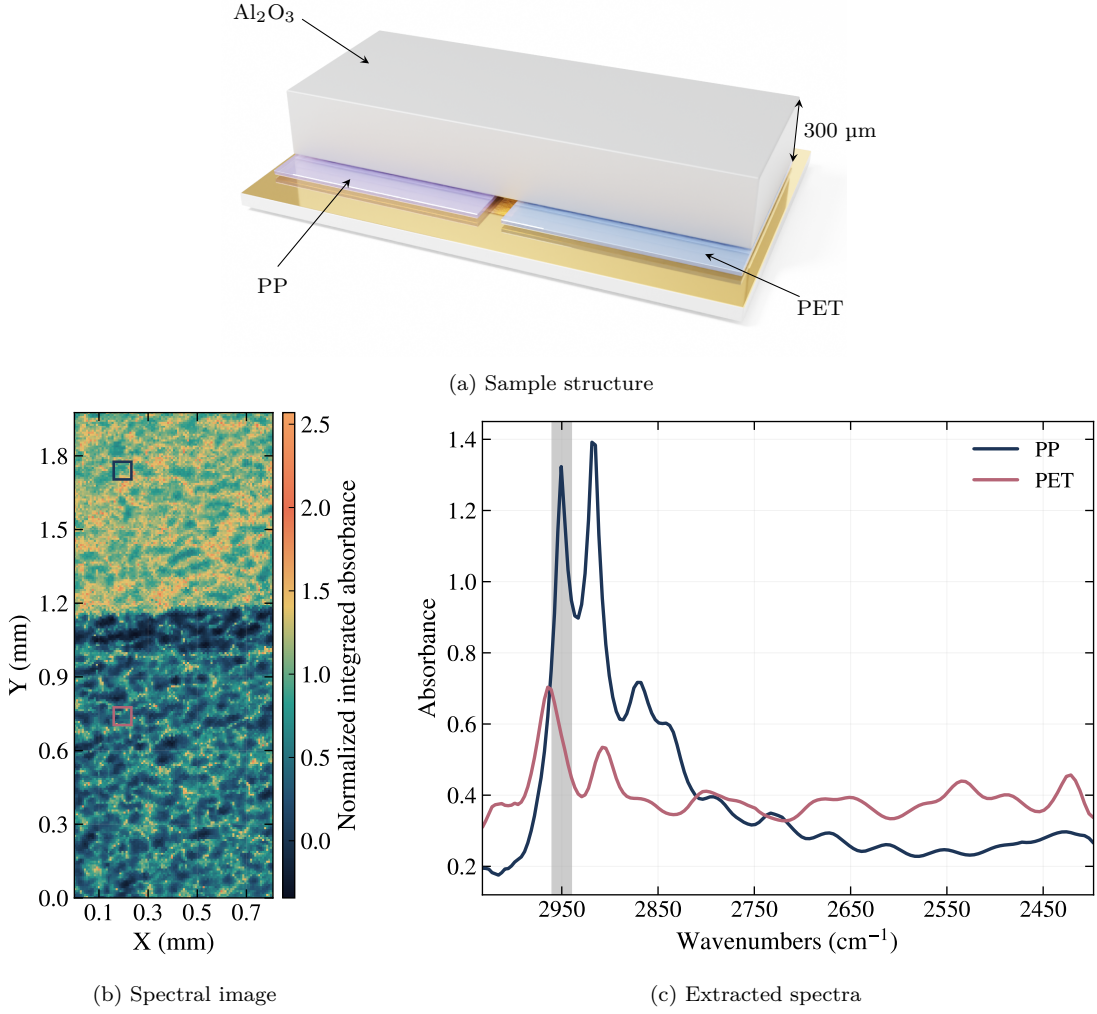

In order to demonstrate the high sensitivity of the system, hyperspectral imaging through a turbid medium was performed. A stacked sample (Fig.~\ref{fig:turbid_sample}) consisting of an alumina (Al$_2$O$_3$, $d=300$~\textmu m) plate on top of two polymer sheets (Polypropylene (PP), $d=4$~\textmu m, and polyethylene terephthalate (PET), $d=25$~\textmu m) and a gold mirror, was investigated. The polished side of the alumina sample was facing up, while the rough side was facing the polymer sheets (see the sample structure in Fig.~\ref{fig:turbid_sample}). A double-sided tape was used to fix the alumina on the gold mirror, thus introducing an air gap between the polymer foils and the alumina. The foils themselves were attached to the gold-coated surface via electrostatic force.
The 247 $\times$ 101 pixel measurement of the sample was performed with an 8~\textmu m spatial step size, a 100~ms integration time per spectrum, and the spectral resolution was set to 27~cm$^{-1}$.  
The background spectrum needed for the calculation of the absorbance was obtained using averaged spectra extracted from the region with a gap between the two polymer foils.

The normalized integrated absorbance image retrieved by integration over the band between 2940--2960~cm$^{-1}$ is shown in Fig.~\ref{fig:turbid_image}. The regions of PP and PET can be clearly differentiated due to the difference in normalized integrated absorbance. Fig.~\ref{fig:turbid_spectra} shows extracted absorbance spectra from the regions indicated with rectangles in the spectral image in Fig.~\ref{fig:turbid_image} and correspond well to spectra from the literature~\cite{Gattinger2025}. PP and PET can be well distinguished via their distinct absorption bands in the range between 3000--2830~cm$^{-1}$. Residual ripples in the spectra, visible especially in the spectral region between 2830--2400~cm$^{-1}$ can be attributed to the speckles and scattering off the 300~\textmu m thick alumina plate on top of the polymers. Therefore, sQFTIR microspectroscopy through a turbid medium at high spatial resolution can provide high-quality spectra that can be unambiguously interpreted and assigned to the respective polymers.

\section{Conclusion}

In this study, we demonstrated the applied capabilities of quantum infrared (IR) spectroscopy for routine chemical analysis of solids. Thus, a scanless quantum FTIR (sQFTIR) microspectrometer and hyperspectral imager that leverages sensing with non-degenerate photon pairs (generated in periodically poled Potassium Titanyl Phosphate) in a nonlinear interferometric arrangement was demonstrated. The system was showcased for analysis of relevant samples such as active pharmaceutical ingredients on metal interfaces; hyperspectral sQFTIR imaging of banknote security features (integrated volumetric); and spectroscopy through turbid media (identification of polymers).
The obtained results delivered high-quality single-pixel spectra (functional group region, 3000~cm$^{-1}$ -- 2400~cm$^{-1}$) at 100~ms exposure time and spatial resolution on the tens-of-micrometer scale. The demonstrated technique and its practical performances are of a great applied potential for various research and industrial fields, in particular accounting for cost-effectiveness and sensitivity of the method compared to state-of-the-art mid-IR systems.

\section*{Funding}
\noindent This project was co-financed by research subsidies granted by the Government of Upper Austria under the QUICK (Wi-2022-597365/18-Au) and QUANTAN (Wi-2026-67023/3-FrJ) projects. The authors acknowledge funding and support from Österreichische Forschungsförderungsgesellschaft (FFG) under the projects QMIRACT (929209) and HAIQUAM (940169).

\section*{Acknowledgment}
\noindent We would like to thank Sven Ramelow, Gregor Langer and Bettina Heise for fruitful discussions.

\section*{Disclosures}

\noindent The authors declare no conflicts of interest.

\section*{Data availability} Data underlying the results presented in this paper are not publicly available at this time but may be obtained from the authors upon reasonable request.

\bibliography{main}

\end{document}